\documentclass[acmsmall]{acmart}
\AtBeginDocument{%
  }

\setcopyright{none}

\acmJournal{JACM}
\acmVolume{37}
\acmNumber{4}
\acmArticle{111}
\acmMonth{8}

\usepackage{cleveref}
\usepackage{enumitem}
\usepackage{multirow}
\usepackage[ruled,vlined,lined,commentsnumbered]{algorithm2e}
\usepackage{booktabs}
\usepackage{moreverb}
\usepackage{fontenc}
\usepackage{amsmath}

\usepackage{amssymb}
\usepackage{fancybox}
\usepackage{color}
\usepackage{colortbl}
\usepackage{array}
\usepackage{diagbox}

\usepackage{flushend}
\usepackage{multicol}
\usepackage{listings}
\usepackage{makecell}
\usepackage{graphicx}
\usepackage{setspace}
\usepackage{soul}
\usepackage{csvsimple}
\usepackage{longtable}
\usepackage[skip=1pt,labelfont=bf]{caption}
\usepackage{url}
\usepackage{lscape}
\usepackage{rotating}
\usepackage{tikz}
\usepackage{enumitem}
\usepackage{subcaption}
\usepackage{wrapfig}
\usepackage[most]{tcolorbox}
\usepackage{hyperref}
\usepackage{pythonhighlight}
\usepackage[normalem]{ulem}
\usepackage{alltt}
\usepackage{tabularray}
\usepackage{balance}
\usepackage{tcolorbox}
\usepackage{xcolor}

\usepackage{titlesec}
\newcommand{\su}[1]{\textcolor{black}{#1}}
\newcommand{\suu}[1]{\textcolor{black}{#1}}

\newcommand{\name}{\textsc{Healer}\xspace}

\begin{document}

\title{Towards Agentic Runtime Healing}

\author{Zhensu Sun}
\affiliation{%
  \institution{Singapore Management University}
  \country{Singapore}
}
\email{zssun@smu.edu.sg}

\author{Haotian Zhu}
\affiliation{%
\institution{Singapore Management University}
\country{Singapore}
}
\email{htzhu@smu.edu.sg}
\author{Bowen Xu}
\affiliation{%
  \institution{North Carolina State University}
  \country{USA}
}
\email{bxu22@ncsu.edu}

\author{Xiaoning Du}
\affiliation{%
  \institution{Monash University}
  \country{Australia}
}
\authornote{Corresponding author.}
\email{xiaoning.du@monash.edu}

\author{Li Li}
\affiliation{%
  \institution{Beihang University}
  \country{China}
}

\email{lilicoding@ieee.org}
\author{David Lo}
\affiliation{%
\institution{Singapore Management University}
\country{Singapore}
}
\email{davidlo@smu.edu.sg}

\begin{abstract}
Self-healing systems have long been a focus of research, aiming to enable software to recover from unexpected runtime errors without human intervention.
Traditional approaches rely on predefined heuristic rules, such as reusing error handlers or rolling back to checkpoints, but these methods struggle to adapt to the diverse range of runtime errors.
The emergence of Large Language Models offers a new opportunity to address this challenge.
Leveraging their ability to understand and generate code and natural language, we propose using LLMs to dynamically generate error-handling strategies in real time, tailored to specific runtime contexts such as error messages and program states.

We demonstrate the feasibility of this approach by designing such a framework, \name, and empirically showing that it can handle runtime errors with a high success rate.
When an unanticipated runtime error occurs, \name leverages its internal LLM to generate bespoke error-handling code.
The LLM is prompted with runtime information, including the error message, error location, and current program state.
The generated healing code is then executed to produce a corrected program state, allowing the program to continue execution with minimal disruption.
We evaluate \name across four code datasets and three state-of-the-art LLMs (GPT-3.5, GPT-4, and CodeQwen-7B), where GPT-4 can successfully recover from 72.8\% of runtime errors, underscoring the promise of LLMs in this domain.
Despite these promising results, challenges remain, particularly regarding the trustworthiness of LLM-generated code and its integration into existing systems.
We mention potential solutions, such as safety checks and \name-aware programming, to mitigate risks and ensure reliable operation.
This work represents the first step toward agentic runtime healing, paving the way for more adaptive, resilient, and self-healing software systems.
\end{abstract}

\ccsdesc[500]{Software and its engineering~Maintaining software}
\ccsdesc[300]{Computing methodologies~Natural language processing}
\ccsdesc[500]{Software and its engineering~Software maintenance tools}

\maketitle

\section{Introduction}\label{sec:intro}

Runtime error handling~\cite{weimer2004finding} is a critical aspect of software reliability engineering.
It requires developers to anticipate potential errors and implement corresponding error handlers in advance. Unanticipated runtime errors, which lack corresponding handlers and are naturally uncaught, can lead to severe consequences such as unexpected program crashes, exposure of sensitive information, and other critical issues~\cite{cwe248}.
To mitigate such risks, significant efforts have been made to identify potential errors in source code through techniques like software testing and verification~\cite{kochhar2015understanding, zhou2012should}.
However, identifying all possible runtime errors and providing corresponding exception handlers remains a challenging and often impractical task.

Given the near inevitability of unanticipated runtime errors, dynamic countermeasures are essential to minimize their impact when they occur.
This necessity has driven the development of self-healing systems~\cite{ghosh2007self, psaier2011survey}, a long-standing research area focused on enabling software to automatically recover from abnormal states and restore functionality.
Importantly, healing a program execution does not involve patching the code to fix an error but rather revising the runtime state, such as variable values and environment configurations, to recover the program's execution.
A successful healing process should not only prevent unexpected termination but also correct the runtime state so that the program can continue meaningfully.
The core challenge for self-healing systems lies in the fact that specific errors and their locations are unknown in advance, requiring real-time responses.
Consequently, these systems rely on predefined heuristic strategies, such as restarting~\cite{candea2003jagr}, rolling back to a checkpoint~\cite{carzaniga2013automatic}, trying alternative methods~\cite{carzaniga2010automatic}, modifying input~\cite{long2012automatic}, reusing existing error handlers~\cite{gu2016automatic}, preparing default error handlers~\cite{rinard2004enhancing}, or combining multiple strategies~\cite{chang2013exception}.
Despite decades of exploration, the reliance on rule-based strategies limits the adaptability of self-healing systems in handling the diverse range of runtime errors that can occur.
The complexity of defining adaptation rules further complicates their implementation in real-world applications.
A large-scale industry study highlights this challenge, emphasizing the need for more flexible and adaptive approaches to self-healing~\cite{weyns2022preliminary}.

\begin{figure*}[t]
    \centerline{\includegraphics[width=\linewidth]{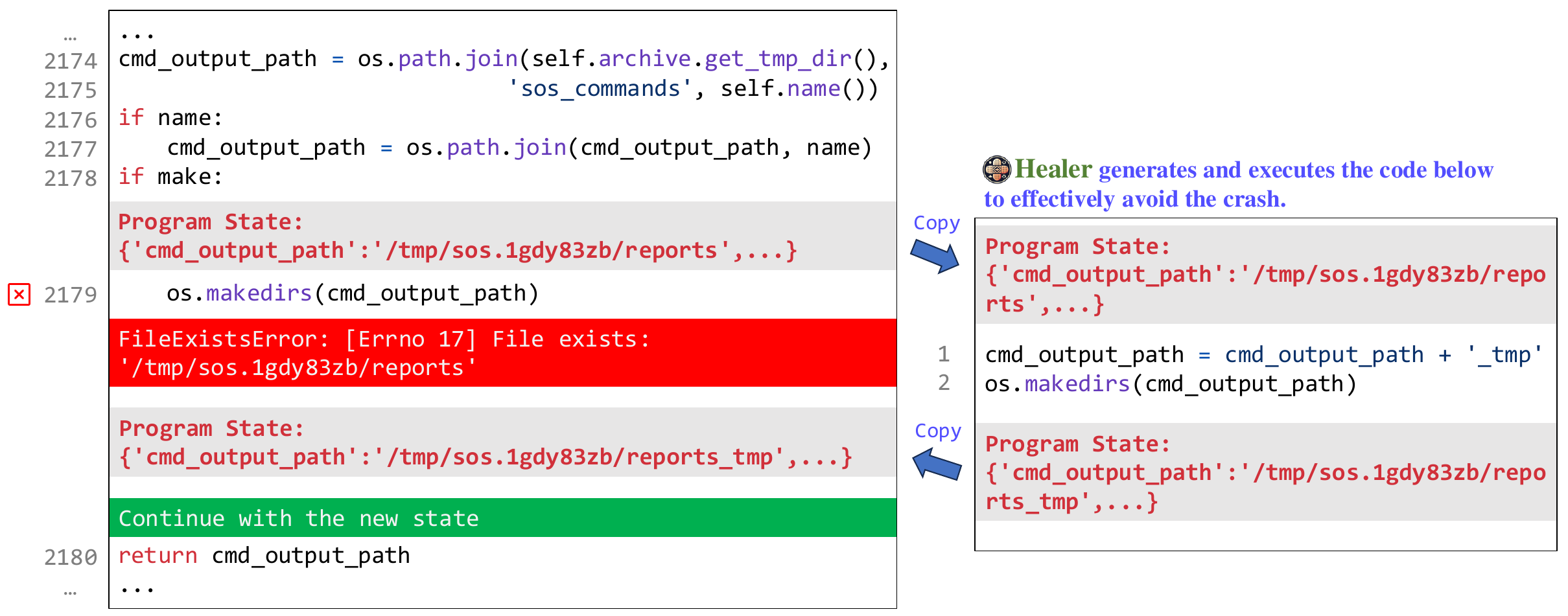}}
    \caption{A motivating example demonstrating how \name handles runtime errors to recover program execution. The code snippet and issue are extracted from SOS~\cite{sos}, an open-source tool for collecting system logs.}
    \label{fig:examples}
\end{figure*}

Recently, the emergence of Large Language Models (LLMs) has introduced new opportunities to revolutionize this field.
LLMs have demonstrated exceptional performance in code-related tasks~\cite{codellm_survey}, such as the AlphaCode2 model~\cite{alphacode2}, which outperformed 85\% of human participants in a programming competition.
These models possess the ability to understand both source code and natural language and can generate a piece of source code within seconds.
Such ability aligns well with the requirements of runtime error handling, i.e., comprehending runtime error contexts, such as error messages and program states, and generating tailored solutions for unanticipated errors in real time.
Specifically, when an unanticipated runtime error occurs, an LLM can be invoked with the runtime error contexts in the prompt and generate an appropriate code script.
This script is not merged into the original program but is executed dynamically to recover the program from the faulty state.
Thus, we propose that using LLMs as runtime error handlers could be a promising pathway toward achieving the long-standing goal of creating truly adaptive and self-healing software systems.

In this paper, we take the first step toward this ambitious goal by demonstrating its feasibility.
Specifically, we design a framework, \name, for LLM-driven runtime error handling and experimentally evaluate its effectiveness.
To illustrate how \name facilitates runtime error handling, we present a motivating example from an open-source tool for collecting system logs, SOS~\cite{sos} (v4.5.5), in~\Cref{fig:examples}.
The extracted code is designed to create a temporary directory to save collected logs. However, when executed in a race condition where \texttt{os.makedirs()} is called after the directory has already been created, the code crashes at Line 2179, resulting in the loss of collected logs.
When such unhandled runtime errors occur, \name can be invoked to handle the error.
It gathers relevant information and instructs its LLM to generate a healing code snippet that seeks to avoid termination caused by the error and enables the program to continue execution with a corrected state.
For this example, the healing code renames the output directory to a new name, avoiding the conflict with the existing directory.
Once generated, the healing code is dynamically executed to produce a new program state, allowing the program to continue execution with the updated state.

Following the design of \name, we implement a prototype for Python and evaluate its effectiveness.
The evaluation is conducted on four different code datasets, containing a total of 1,185 (code, test case) pairs, where the code raises runtime errors with the test case.
We test \name empowered by three state-of-the-art LLMs: two closed-source LLMs, GPT-3.5 and GPT-4, and one open-source LLM, CodeQwen.
Initially, we investigate whether \name, under a zero-shot setting, could effectively avoid interruption and correct faulty program states for runtime errors during execution.
The results show that without any fine-tuning, GPT-4 could continue execution in 72.8\% of the instances and produce correct results in 39.6\% of the instances.
This demonstrates that LLMs, even without sophisticated measures, are capable of correctly handling runtime errors, revealing their great potential for this task.
Additionally, we create an instruction-tuning dataset using the (prompt, healing code) pairs generated in our experiments and fine-tune GPT-3.5 and CodeQwen to assess the effectiveness of fine-tuning.
After fine-tuning, the performance of GPT-3.5 and CodeQwen improves by 31.7\% and 23.1\%, respectively, in instances free of interruption, and by 26.8\% and 65.8\% in instances yielding correct results.
Notably, the fine-tuned GPT-3.5 achieves performance comparable to GPT-4. 
We thus conclude that using LLMs for runtime error handling is a promising direction with great potential.

Although \name has shown promising results, there are remaining challenges that must be addressed before LLMs can be applied in practice.
The foremost challenge is the trustworthiness of LLM-generated code, as the generated code is executed directly, and LLMs may produce incorrect results, introducing potential risks to the program.
To better understand this trustworthiness issue, we empirically investigate the healing code generated by GPT-4 and observe some risky behaviors, such as ignoring key logic in the original code, which could lead to unexpected program behaviors.
Therefore, how to mitigate potential risks from the black box of LLMs is a challenging and important problem.
Fortunately, there are many promising solutions to mitigate these risks.
For example, we can predefine a set of "safe" functions, such as a function for data saving, and only allow the LLM to use these functions to perform the healing.
With such measures, we may not be far from achieving a fully self-healing system.

\section{Proposed Framework}\label{sec:method}

\begin{figure}[t]
    \centerline{\includegraphics[width=0.8\linewidth]{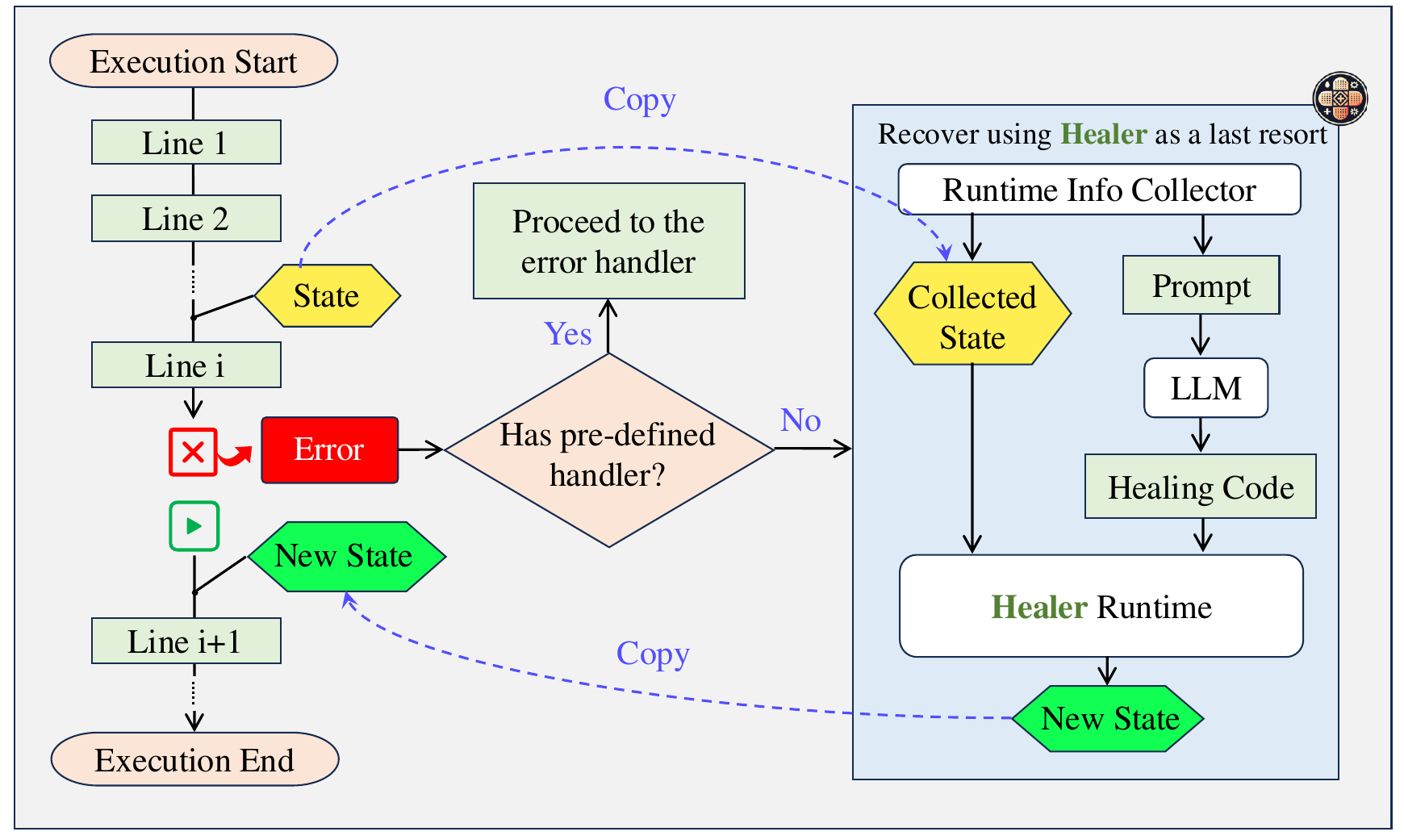}}
    \caption{The workflow of \name. When a runtime error occurs and no pre-defined error handler is available, \name is activated. It collects the error context, prompts the LLM, and generates healing code. The healing code is then executed in \name's runtime environment to produce a new program state, allowing the program to continue execution.}
    \label{fig:framework}
\end{figure}
 
\name handles the errors relying on an integrated LLM, which generates a piece of code, named healing code, by understanding the error context.
\su{The integrated LLM refers to a model that is logically incorporated into the automated runtime error handling loop, regardless of its physical deployment. It enables the use of diverse backend models (either local or remote) based on computing resource availability.}
The workflow of \name is illustrated in~\Cref{fig:framework}.
\su{To ease the understanding, we assume the framework is applied for an interpreted programming language, such as Python and JavaScript, where the code is executed line by line, and the runtime environment automatically persists program state between each execution step.}
Formally, consider a program $P$ with $n$ lines, represented as $P = \{L_1, L_2, ..., L_n\}$.
If an error $E$ occurs at line $L_i$ without an available error handler, \name is activated.
\name first collects the error context, including the error message, the program state $S$, the entire code snippet, and the erroneous line, to construct a prompt.
The integrated LLM is then instructed with this prompt to generate a piece of healing code $L'_i$.
This code $L'_i$ is executed under the state $S$ of the original program.
If the execution of $L'_i$ is successful, the resulting program state $S'$ is updated back to the original program state, allowing the program to continue from line $L_{i+1}$. 
Notably, while the state is corrected for the current session, the original program is not permanently fixed and the bug will still need to be reproduced and fixed in future development.
If the execution of the healing code still raises errors, the program is terminated, and the error is reported, following traditional error-handling mechanisms.

\begin{figure*}[t]
    \centerline{\includegraphics[width=\linewidth]{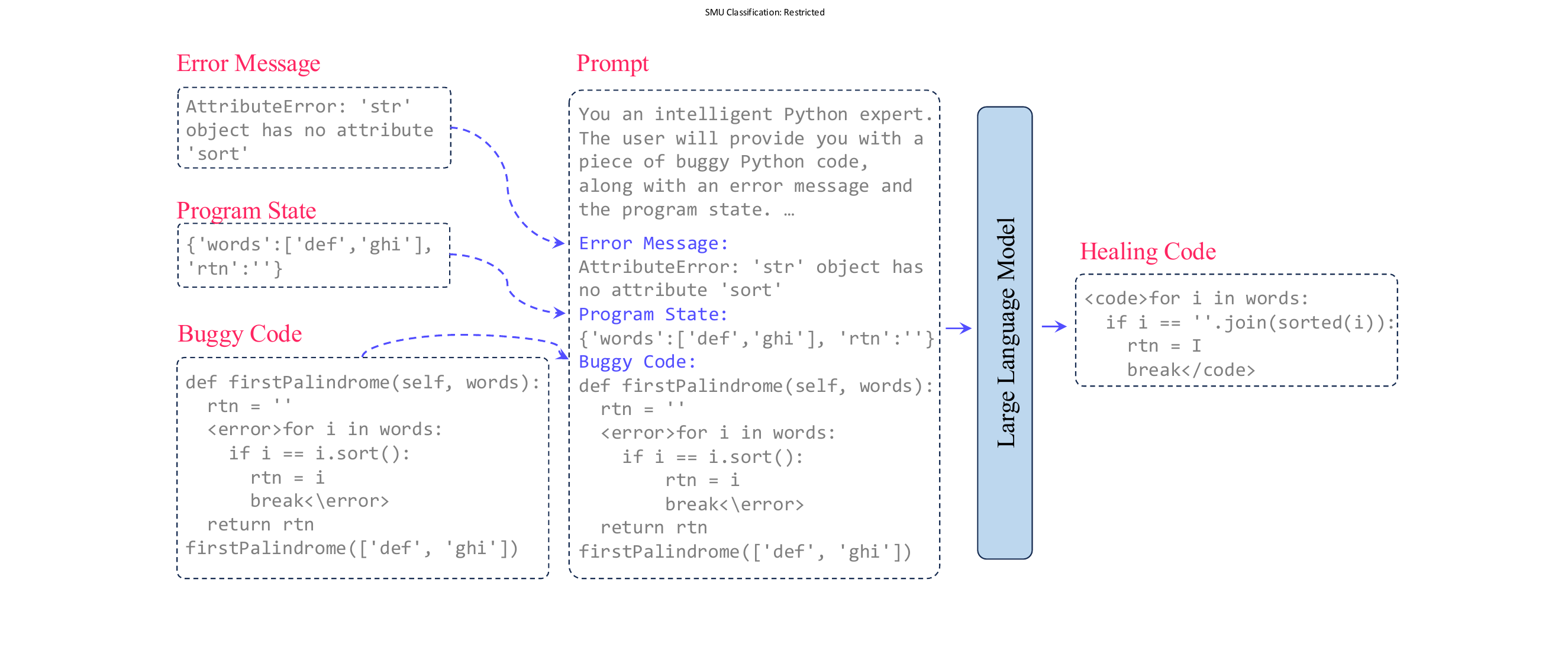}}
    \caption{An example of the prompt construction workflow in \name. \su{The prompt is constructed based on a pre-defined template.}}
    \label{fig:prompt}
\end{figure*}

The \name framework consists of three main components:
\begin{itemize}[leftmargin=*]
\item \textbf{Runtime Information Collector.} Once \name is activated, the runtime information collector gathers necessary information from the runtime environment.
Specifically, \name collects four types of information: the source code, error message, error location, and program state before the error is raised.
This information is essential for a human developer to understand the error and write the healing code, and thus \name collects it to prompt the LLM effectively.

\item \textbf{Large Language Model.}
\su{As illustrated in~\Cref{fig:prompt}, \name interacts with the LLM using a structured prompt template that combines task instructions with relevant context fields. The instructions outline the task description, goals, and constraints, supplemented by an example to guide the LLM's generation. Following this, the template is populated with the collected error context; specifically, the entire source code, error message, and program state are inserted into their respective fields. Adhering to OpenAI's best practices for prompt engineering~\cite{openai_prompt_engineering_delimiters}, each field is clearly demarcated by a section title, including Error Message:'', Program State:'', and ``Buggy Code:''. Notably, the error location is not designated as a standalone section. Instead, we highlight the error-raising logic by wrapping the error-raising block within `<error>' and `</error>' tags directly in the source code. Due to space limitations, the full prompt is provided in our artifact.}

\item \textbf{Healing Runtime.}
The healing code generated by the LLM needs to be executed to produce a new program state.
We refer to the runtime that executes the healing code as the healing runtime.
A straightforward approach is to execute the healing code in the same runtime as the original program.
However, to mitigate potential security risks associated with LLM-generated code, alternative runtime environments can be used.
For example, the healing runtime could be a new session, a sandbox, or a virtual machine, depending on the required security level.
Notably, the execution does not re-run the original program from the beginning but rather executes the healing code with the program state that existed before the error occurred.
\end{itemize}

\section{Evaluation of Healer}\label{sec:results}
We further evaluate the effectiveness of \name in handling runtime errors. To conduct this evaluation, we implement a proof-of-concept Python code executor that accepts source code strings as input and executes them using the \name process.

\begin{table*}[t]
    \caption{The performance of GPT-4, GPT-3.5, and CodeQwen on all the benchmarks in a zero-shot setting. \su{Benchmarks with model names in parentheses were generated by that specific LLM. CORR., PROC. and FAIL. respectively refer to the percentage of healing results where the program produces correct results, proceeds with execution, or still raises errors}}
    \centering
    \scalebox{0.9}{\begin{tblr}{
  cells = {c},
  colsep = 1.5pt,
  cell{1}{1} = {r=2}{}, 
  cell{1}{2} = {r=2}{}, 
  cell{1}{3} = {c=3}{}, 
  cell{1}{6} = {c=3}{}, 
  cell{1}{9} = {c=3}{}, 
  vlines,
  hline{1,3,4,10,16-18} = {-}{},
  hline{2} = {3-11}{},
}
\textbf{Benchmark} & \textbf{\#Errors} & \textbf{GPT-4} & & & \textbf{GPT-3.5} & & & \textbf{CodeQwen} & & \\
 & & \textbf{CORR.} & \textbf{PROC.} & \textbf{FAIL.} & \textbf{CORR.} & \textbf{PROC.} & \textbf{FAIL.} & \textbf{CORR.} & \textbf{PROC.} & \textbf{FAIL.}\\
CodeNet & 228 & 42.1\% & 73.2\% & 26.8\% & 31.1\% & 52.6\% & 47.4\% & 16.7\% & 39.9\% & 60.1\% \\
HumanEval (CodeLlama) & 70 & 47.1\% & 62.9\% & 37.1\% & 35.7\% & 57.1\% & 42.9\% & 12.9\% & 32.9\% & 67.1\% \\
HumanEval (CodeT5+) & 57 & 36.8\% & 64.9\% & 35.1\% & 21.1\% & 50.9\% & 49.1\% & 14.0\% & 36.8\% & 63.2\% \\
HumanEval (GPT-3.5) & 43 & 23.3\% & 72.1\% & 27.9\% & 7.0\% & 37.2\% & 62.8\% & 14.0\% & 46.5\% & 53.5\% \\
HumanEval (GPT-4) & 19 & 57.9\% & 100.0\% & 0.0\% & 21.1\% & 63.2\% & 36.8\% & 10.5\% & 21.1\% & 78.9\% \\
HumanEval (Mistral) & 37 & 35.1\% & 64.9\% & 35.1\% & 18.9\% & 45.9\% & 54.1\% & 10.8\% & 27.0\% & 73.0\% \\
HumanEval (StarCoder) & 69 & 37.7\% & 58.0\% & 42.0\% & 30.4\% & 39.1\% & 60.9\% & 18.8\% & 34.8\% & 65.2\% \\
MBPP (CodeLlama) & 30 & 40.0\% & 80.0\% & 20.0\% & 26.7\% & 50.0\% & 50.0\% & 13.3\% & 30.0\% & 70.0\% \\
MBPP (CodeT5+) & 49 & 44.9\% & 75.5\% & 24.5\% & 16.3\% & 38.8\% & 61.2\% & 10.2\% & 40.8\% & 59.2\% \\
MBPP (GPT-3.5) & 14 & 7.1\% & 92.9\% & 7.1\% & 0.0\% & 50.0\% & 50.0\% & 0.0\% & 28.6\% & 71.4\% \\
MBPP (GPT-4) & 3 & 33.3\% & 100.0\% & 0.0\% & 33.3\% & 33.3\% & 66.7\% & 0.0\% & 0.0\% & 100.0\% \\
MBPP (Mistral) & 73 & 34.2\% & 68.5\% & 31.5\% & 13.7\% & 34.2\% & 65.8\% & 11.0\% & 34.2\% & 65.8\% \\
MBPP (StarCoder) & 57 & 36.8\% & 63.2\% & 36.8\% & 29.8\% & 49.1\% & 50.9\% & 12.3\% & 42.1\% & 57.9\% \\
LiveCodeBench & 47 & 17.0\% & 76.6\% & 23.4\% & 10.6\% & 51.0\% & 49.0\% & 6.4\% & 29.8\% & 70.2\% \\
DebugBench & 436 & 40.6\% & 77.5\% & 22.5\% & 22.2\% & 49.3\% & 50.7\% & 12.2\% & 33.0\% & 67.0\% \\
\end{tblr}}
    \label{tab:rq1}
\end{table*}

\subsection{LLMs Can Effectively Handle Runtime Errors}
\label{sec:rq1}

\su{We evaluate \name on four benchmarks: \textbf{CodeNet}~\cite{IBMProjectCodeNet} for human-written code, \textbf{HumanEval+}, \textbf{MBPP+}~\cite{liu2024your}, and \textbf{LiveCodeBench}~\cite{jain2024livecodebench0} for LLM-generated code, and \textbf{DebugBench}~\cite{tian2024debugbench} for code with implanted bugs.}
For CodeNet and DebugBench, we execute the code samples with the provided test cases and retain the (code, test case) pairs that raise runtime errors, yielding 228 and 436 instances, respectively.
In the paper of \textbf{HumanEval+} and \textbf{MBPP+}, multiple LLMs are evaluated, and the code snippets generated during the evaluation are publicly available~\cite{EvalPlusReleases}.
We adopt the code snippets generated by six different LLMs, including GPT-3.5 (ChatGPT)~\cite{openai_gpt3_5_turbo}, GPT-4 (gpt-4-1106-preview)~\cite{achiam2023gpt}, CodeLlama (34B)~\cite{roziere2023code}, CodeT5+ (16B)~\cite{wang2023codet5}, Mistral (7B)~\cite{jiang2023mistral}, and StarCoder (15B)~\cite{li2023starcoder}, and then follow the same execution-then-selection process to obtain the instances.
Finally, we obtain 295 instances from HumanEval+ and 226 instances from MBPP+.
\su{Similarly, we collect the error-raising code snippets generated by GPT-4 (gpt-4-1106-preview) from LiveCodeBench as the control group that is free from potential data leakage issue, resulting in 47 samples.}
For each instance, we execute its code snippet with the paired test case and handle the runtime errors using the three LLMs, GPT-3.5, GPT-4, and CodeQwen, following the framework described in~\Cref{sec:method}.
\su{We record the final execution status of each instance and compute the three metrics: CORRECT, PROCEED, and FAIL, respectively indicating whether the program finally produces correct results expected by the test case, whether the program can continue its execution, and whether the healing code still raise errors.}

The results, shown in~\Cref{tab:rq1}, demonstrate that \name can effectively heal runtime errors, thereby improving program availability.
\su{On average across all benchmarks, GPT-4, GPT-3.5, and CodeQwen can proceed with 73.0\%, 48.3\%, and 35.1\% of of the executions, respectively, and help 38.7\%, 23.5\%, and 13.0\% of the programs produce correct results.}
This provides compelling evidence that using LLMs to handle runtime errors is a feasible and promising direction.
Among the LLMs, GPT-4 consistently achieves the best performance across all benchmarks, with the highest CORRECT and PROCEED metrics.
For example, on the DebugBench benchmark, GPT-4 successfully completes the execution of 338 instances and corrects the faulty states for at least 177 instances, accounting for 77.5\% and 40.6\% of the total errors.
Even for code snippets generated by LLMs, such as those in \su{HumanEval, MBPP, and LiveCodeBench}, \name improves the execution outcomes.
For instance, GPT-4 enhances the execution results of the code it generates by successfully correcting 11 additional instances in HumanEval and 1 additional instance in MBPP.
\su{Moreover, \name achieves comparable performance on LiveCodeBench, demonstrating the negligible impact of potential data leakage issues.}
This illustrates that LLMs can be used to mitigate the runtime errors caused in their own-generated code.

\suu{While we primarily focus on comparing different LLMs, it is worth contextualizing these results against traditional self-healing strategies, such as Restarting, Retrying, or Steamrolling (i.e. logging the runtime error and continuing execution despite it).
In the context of our benchmarks, the runtime errors are deterministic errors (e.g., IndexError, AttributeError, NameError), instead of environmental failures.
Consequently, traditional strategies like Retry or Rollback would yield a 0\% success rate in proceeding execution, as re-executing the same erroneous logic without adaptive healing leads to the same crash.
Similarly, a error steamroller~\cite{fuckitpy,fuckitjs} might allow the program to survive, thereby improving the PROCEED metric.
However, it cannot correct the program state to complete the task, resulting in an extremely low CORRECT rate.
For example, when we adopt a error steamroller that exceed the code execution anyway, the CORRECT metric is only 4.3\%.
This underscores HEALER's unique value in bridging the gap between simple crash avoidance and functional recovery.}

\textbf{Takeaway \#1:}
\name can effectively handle runtime errors and improve the availability and reliability of programs.
Notably, GPT-4-based \name can continue execution in 73.0\% of the instances and produce correct results in 38.7\% of the instances.

\subsection{Fine-Tuning Further Improves Healer's Effectiveness}

\begin{table}[t]
    \centering
    \caption{The comparison between fine-tuned and original versions for GPT-3.5 and CodeQwen. GPT-4 is used as a reference as we don't have access to its fine-tuning API.}
    
    \scalebox{0.9}{\begin{tblr}{
  cells = {c},
  cell{1}{1} = {r=2}{},
  cell{1}{2} = {r=2}{},
  cell{1}{3} = {c=4}{},
  cell{2}{3} = {c=2}{},
  cell{2}{5} = {c=2}{},
  cell{4}{1} = {r=2}{},
  cell{6}{1} = {r=2}{},
  vline{1-3,5,7} = {-}{},
  hline{1,3-4,6,8} = {-}{},
  hline{2} = {3-6}{},
}
\textbf{LLM} & \textbf{Version} & \textbf{CodeNet (228)} &  &  & \\
 &  & \textbf{CORRECT} &  & \textbf{PROCEED} & \\
GPT-4 & Original & 96 &  & 167 & \\
GPT-3.5 & Original & 71 &  & 120 & \\
 & Finetuned & \textbf{90} & 26.8\%\textuparrow & \textbf{158} & 31.7\%\textuparrow\\
CodeQwen & Original & 38 &  & 91 & \\
 & Finetuned & \textbf{63} & 65.8\%\textuparrow & \textbf{112} & 23.1\%\textuparrow
\end{tblr}}
    
    \label{tab:rq3}
\end{table}

Using the data points that produce correct results in the previous experiment, we create an instruction-tuning dataset to fine-tune the LLMs for runtime error handling.
This dataset consists of 586 (prompt, healing code) pairs, generated from samples across all benchmarks except the human-written benchmark, CodeNet, which is reserved as the test set for the fine-tuned LLMs.
Specifically, we run GPT-3.5 and GPT-4 to handle the runtime errors in each instance from the selected benchmarks and record the healing code.
If the healing result is CORRECT, we pair the prompt with its corresponding healing code as a sample in the dataset.
With this dataset, we fine-tune two LLMs, GPT-3.5 and CodeQwen, while using GPT-4's performance as a reference.
\su{Specifically, GPT-3.5 is fine-tuned via the commercial API provided by OpenAI, where the hyper-parameters follow its default settings. CodeQwen is fine-tuned on our machine using LoRA (Low-Rank Adaptation), a parameter-efficient fine-tuning technique~\cite{hu2021lora}, with its hyperparameters available in our artifacts.}
After fine-tuning, we evaluate the fine-tuned LLMs on CodeNet.

The results, presented in~\Cref{tab:rq3}, show that fine-tuning effectively enhances the LLMs' performance in handling runtime errors.
Both GPT-3.5 and CodeQwen exhibit substantial improvements, with GPT-3.5 achieving 26.8\% more CORRECT instances and 31.7\% more PROCEED instances, and CodeQwen achieving 65.8\% more CORRECT instances and 23.1\% more PROCEED instances.
Remarkably, the fine-tuned GPT-3.5 successfully produces 90 CORRECT instances, closely approaching GPT-4's performance of 96 CORRECT instances.
Similarly, the fine-tuned CodeQwen also achieves competitive results comparable to GPT-3.5.
This indicates that fine-tuning can be a powerful approach to developing task-specific LLMs for runtime error handling, potentially surpassing general-purpose LLMs in effectiveness.
Therefore, a domain-specific LLM for runtime error handling is a promising direction.

{\textbf{Takeaway \#2:}} Fine-tuning significantly improves LLM performance in handling runtime errors. GPT-3.5 demonstrates a 26.8\% increase in CORRECT instances and a 31.7\% increase in PROCEED instances on CodeNet, achieving performance comparable to GPT-4.

\section{Remaining Challenges}

\begin{table}[t]
    \centering
    \caption{A summary of the behaviors observed in the healing code generated by GPT-4.}
    \scalebox{0.9}{\begin{tblr}{
  row{1} = {c},
  row{11} = {c},
  cell{2}{2} = {c},
  cell{3}{2} = {c},
  cell{4}{2} = {c},
  cell{5}{2} = {c},
  cell{6}{2} = {c},
  cell{7}{2} = {c},
  cell{8}{2} = {c},
  cell{9}{2} = {c},
  cell{10}{2} = {c},
  cell{12}{2} = {c},
  cell{13}{2} = {c},
  cell{14}{2} = {c},
  cell{15}{2} = {c},
  cell{16}{2} = {c},
  cell{17}{2} = {c},
  cell{18}{2} = {c},
  cell{19}{2} = {c},
  cell{20}{2} = {c},
  cell{21}{2} = {c},
  vlines,
  hline{1-2,11-12,22} = {-}{},
}
\textbf{Recognizable behaviors~ (consistent patterns identifiable by static analysis tools)} & \textbf{Count} \\
\textbf{No Action}: Repeat the code without any semantical changes                                     & 1 (0.5\%)      \\
\textbf{Value Change}: Assign/define a new value for a variable/constant                               & 43 (21.0\%)    \\
\textbf{Value Change (IF)}: Conditionally assign a new value for a variable                            & 13 (6.3\%)     \\
\textbf{Variable Initialization}: Initialize an undefined variable                                     & 21 (10.2\%)    \\
\textbf{Type Change}: Set/modify the type of a variable/constant                                       & 12 (5.9\%)     \\
\textbf{Library Import}: Import a new library/API                                                      & 56 (27.3\%)    \\
\textbf{Function Invocation}: Invoke an existing custom function with new arguments                    & 3 (1.5\%)      \\
\textbf{Variable Change}: Change the reference to a different variable                                 & 20 (9.8\%)     \\
\textbf{Operator Change}: Use a different operator in a statement or expression                        & 7 (3.4\%)      \\
\textbf{Unrecognizable behaviors~ (irregular or unpredictable patterns)}                               & \textbf{Count} \\
\textbf{Function Defination}: Define a new function and utilize it in the logic                        & 1 (0.5\%)      \\
\textbf{Function Replacement}: Implement the logic as an alternative to a function call                & 8 (3.9\%)      \\
\textbf{Type Change\&Usage}: Set a new type to a variable and utilize it in the logic                  & 2 (1.0\%)      \\
\textbf{API Replacement}: Invoke an alternative API                                                    & 12 (5.9\%)     \\
\textbf{Indexing Omission}: Omit the indexing to a variable                                            & 1 (0.5\%)      \\
\textbf{Library Import\&Usage}: Import a new library and utilize it in the logic                       & 11 (5.4\%)     \\
\textbf{Context Omission}: Skip the remaining logic in error-raising code, except for specific lines   & 5 (2.4\%)      \\
\textbf{Return Omission}: Omit a return statement                                                      & 6 (2.9\%)      \\
\textbf{Context Duplication}: Repeat the entire given code context                                     & 1 (0.5\%)      \\
\textbf{Function Call Omission}: Omit the invocation of an undefined function                          & 3 (1.5\%)      
\end{tblr}}
    \label{tab:cases}
\end{table}

\name operates autonomously, executing code generated by LLMs to handle runtime errors without human intervention.
However, research has shown that LLM-generated code can exhibit unintended behaviors, introducing risks when executed~\cite{pearce2022asleep, yang2024robustness}.
This issue has also been highlighted in other studies that execute model-generated code directly, such as LExecutor~\cite{souza2023lexecutor}, raising concerns about the reliability of \name.
For instance, one might ask, "What if the LLM deletes my database?".

\subsection{LLMs can introduce unexpected behaviors}\label{sec:investigation}
To better understand the risks, we collect all the incorrect healing code generated by GPT-4 on the DebugBench, i.e., the healing code that successfully proceeds the execution but finally produces incorrect results.
This process yields 204 pieces of healing code.
\su{Healing code that fails to execute in the Healer Runtime is excluded from the analysis, since it does not affect the original runtime environment}.
Two authors, each with over three years of experience as Python developers, manually examine these collected instances to identify the behaviors that LLMs might introduce when healing runtime errors.
The goal is to empirically discover the behaviors an LLM could cause during the healing process.
Specifically, the two authors review each healing code together and discuss the behaviors introduced compared to the original error-raising code.
The identified behaviors are incrementally added to a list.
Note that we do not aim to create a comprehensive taxonomy of all possible behaviors but rather to provide a quantitative demonstration of \name's behaviors.

In total, we identify 19 distinct behaviors, detailed in~\Cref{tab:cases}.
Some of these behaviors, such as omitting part of the context (2.4\%) or omitting return statements (2.9\%), are risky and can disrupt the expected flow of data, leading to unexpected execution paths.
Therefore, without further constraints, the recovery performed by \name is not guaranteed to be correct or safe.
Nonetheless, our findings indicate that these risks are manageable because the majority of these behaviors follow distinct patterns that can be precisely identified through rule-based comparisons between the healing code and the error-raising code.
Specifically, 75\% of the investigated samples exhibit recognizable behaviors, such as changing variable values (21.0\%), initializing undefined variables (10.2\%), adjusting types (5.9\%), or importing new libraries (27.3\%).
These behaviors can be identified through static checks based on a predefined whitelist.
In this way, \name can be conservatively configured to perform only behaviors that are not risky to the guarded program.
While these restrictions may limit \name's capabilities, they significantly reduce the uncertainty associated with its operation.

\subsection{A Trustworthy Future?}
Humans also make mistakes.
However, with techniques like software testing, we can identify and mitigate these risks to an acceptable level.
Similarly, in addition to developing more powerful LLMs, supporting mechanisms for \name are crucial for ensuring reliability.
As a newly proposed idea, the ecosystem for \name is currently underdeveloped.
Below, we discuss potential methods to help LLMs behave more reliably.
By implementing these strategies, we can move closer to a future where \name operates reliably and safely within software systems, minimizing risks while maximizing its potential for automated error recovery.

\noindent \textbf{Safety Checks:}
Behaviors such as Context Omission (skipping parts of the original logic) or Return Omission (omitting a return statement) can significantly disrupt the expected flow of the program.
Mechanisms like static checkers can be used to prevent the execution of potentially harmful code.
For example, we can define a whitelist of behaviors that \name is allowed to perform and use static checkers to ensure that the healing code does not contain any behaviors outside this whitelist.
Additionally, techniques used for source code, such as dynamic testing, static analysis, and verification, may also be applicable to the healing code.

\noindent \textbf{\name-aware Programming:}
Traditional software systems are fundamentally deterministic, and the uncertainty introduced by \name can sometimes be unacceptable.
To better accommodate \name, software systems need to be designed with its potential interventions in mind.
For example, since \name may change variable values, the system should include additional mechanisms to handle these changes, such as warning users to be cautious about the system's outputs or maintaining detailed error logs for future bug fixing.
Moreover, it is important to establish a ``comfort zone'' for \name within the system, where its interventions are acceptable and unlikely to cause significant harm.
This comfort zone could encompass non-critical modules or components, such as user interface elements, where the potential side effects of \name's operations are more tolerable.
By confining \name's activities to these safe areas, developers can benefit from automated error recovery without exposing more sensitive parts of the system to unnecessary risk.

\noindent \textbf{Reliable Skillsets for LLMs:}
Inspired by practices in AI agents for software development~\cite{wang2024openhands}, a possible solution is to provide a toolkit as a skill set for the LLM.
The LLM inside \name would then be restricted to using only the functions in this toolkit.
For example, the toolkit could provide a predefined data-saving function, and the LLM could use this function to save data that might otherwise be lost.
Since these functions are predefined, the behaviors of \name become predictable, thereby mitigating risks.

\section{Discussion}\label{sec:discussion}
In this section, we discuss the potential issues with \name, including its deployment implication, and extension to other programming languages.

\noindent\textbf{Infrastructure and Deployment Implications}
\su{Technically, \name requires (1) access to runtime state information (e.g., stack traces and variable states), (2) an execution environment (e.g., a sandbox, container, or lightweight virtual machine) to properly execute LLM-generated healing code, and (3) access to an LLM service, either via a remote API or a locally deployed model. Among these prerequisites, the LLM service imposes additional requirements on the software, where the choice between remote APIs and local deployment involves distinct architectural trade-offs.
Using remote APIs requires stable network connectivity to the server-hosted model. 
Conversely, employing a local model requires the hosting device to possess sufficient computational power to support inference. To address this, various techniques such as model compression~\cite{shi2022compressing} and quantization~\cite{liang2021pruning} have been developed to enable inference on user devices. For example, LLAMA.cpp~\cite{gerganov_llama_cpp} can efficiently run gpt-oss-20B~\cite{gptoss} on a single M3 Ultra MacBook. As the field advances, we anticipate further reductions in the resource requirements for LLMs.
}

\noindent\textbf{\name in Real-world Systems}
\su{To effectively apply \name in real-world applications, some engineering challenges need to be addressed. For instance, real-world software systems can contain millions of lines of code, the majority of which are irrelevant to any single runtime error. Consequently, for LLMs, reading through the entire codebase for every error is inefficient. This necessitates a targeted context retrieval strategy to identify code segments relevant to the error trajectory, methods that are already widely studied in repository-level tasks such as bug localization~\cite{6227210} and repository code completion~\cite{zhang-etal-2023-repocoder}. Furthermore, \name must integrate seamlessly with existing error recovery mechanisms. For example, it should function alongside traditional rollback strategies (e.g., snapshot restoration) so that if \name fails to resolve the error, the system can default to these predefined backup measures.}

\noindent\textbf{Extension to other Programming Languages}
\su{While our current experiments implement \name in Python, the underlying framework for handling runtime errors is designed to be language-agnostic. Interpreted languages (e.g., Python, JavaScript, Ruby) naturally fit \name because they natively support dynamic code execution. In contrast, compiled languages (e.g., Java, C++, Rust) require a compilation step prior to execution, introducing architectural complexity to the dynamic execution process. However, this barrier is not insurmountable through specific language features. For instance, Java provides the Java Compiler API to compile code fragments at runtime, and C++ allows for dynamic loading of shared libraries. While some engineering effort is required to achieve this, extending \name to these languages is promising. We look forward to exploring these extended implementations in our future work.}

\section{Conclusion}
AI-driven runtime error handling represents a promising approach to enhancing the reliability of software systems and could serve as a new pathway toward achieving self-healing systems.
For the first time, we propose this idea and provide a proof-of-concept implementation to demonstrate its feasibility.
Through our implementation, we experimentally show that \name can handle runtime errors with a high success rate.
However, significant challenges remain, particularly regarding the uncertainty of LLM behavior and the compatibility of such systems with existing software architectures.
We hope that this work will inspire further research in this area and pave the way for the future of AI-driven runtime error handling, ultimately leading to more robust and resilient software systems.
The artifact is available at \url{https://github.com/v587su/Healer}.

\section*{Author Bios}
\textbf{Zhensu Sun} is a PhD Student at Singapore Management University, Singapore, Singapore.

\textbf{Haotian Zhu} is a Research Engineer at Singapore Management University, Singapore, Singapore.

\textbf{Bowen Xu} is a Assistant Professor at North Carolina State University, Raleigh, North Carolina, USA.

\textbf{Xiaoning Du} is a Senior Lecturer at Monash University, Melbourne, Victoria, Australia.

\textbf{Li Li} is a Professor at Beihang University, Beijing, China.

\textbf{David Lo} is a Professor at Singapore Management University, Singapore, Singapore.

\bibliographystyle{ACM-Reference-Format}
\bibliography{sample-base}

\end{document}